\documentclass[aps,pra,reprint,twocolumn,showpacs,superscriptaddress,floatfix]{revtex4-1}
\usepackage{amsmath}
\usepackage{amssymb,dsfont}
\usepackage[dvips]{graphicx}
\usepackage[loose]{subfigure}
\newcommand{\imu}{\textrm{i}}

\begin{document}
\title{Entanglement trapping in a non-stationary structured reservoir}
\date{\today}
\author{C.~Lazarou}
\affiliation{Department of Physics and Astronomy, University College London, Gower Street, London WC1E-6BT, UK} 
\author{K. Luoma}
\affiliation{Turku Centre for Quantum Physics, Department of Physics and Astronomy, 
University of Turku, FI-20014 Turun yliopisto, Finland}
\author{S. Maniscalco}
\affiliation{Turku Centre for Quantum Physics, Department of Physics and Astronomy, 
University of Turku, FI-20014 Turun yliopisto, Finland}
\affiliation{SUPA, Department of Physics, Heriot-Watt University, Edinburgh, EH14 4AS, UK}
\author{J. Piilo}
\affiliation{Turku Centre for Quantum Physics, Department of Physics and Astronomy, 
University of Turku, FI-20014 Turun yliopisto, Finland}
\author{B.~M.~Garraway}
\affiliation{Department of Physics and Astronomy, University of Sussex, Falmer, 
Brighton, BN1 9QH United Kingdom}

\begin{abstract}
We study a single two-level atom interacting with a reservoir of modes defined by a reservoir structure function 
with a frequency gap. Using the pseudomodes technique, we derive the main features of a trapping state formed 
in the weak coupling regime. Utilising different entanglement measures we show that strong correlations and entanglement 
between the atom and the modes are in existence when this state is formed. Furthermore, an unexpected feature for the 
reservoir is revealed. In the long time limit and for weak coupling the reservoir spectrum is not constant in time. 
\end{abstract}
\maketitle
\section{Introduction} \label{intro}
In recent years, entanglement and quantum correlations have attracted the attention of many physicists 
working in the area of quantum mechanics \cite{Amico2008,Horodecki2009}. 
This is due to the ongoing research in the area of quantum information \cite{Nielsen},
and also because of the advances made in different experimental disciplines, 
such as in
ion traps \cite{Leibfried2003} and Bose-Einstein condensation \cite{Dalfovo1999,Leggett2001}.
Developments in the field of cavity QED, where experiments in the strong
coupling regime are carried out \cite{Raimond2001,Varcoe2004} provide plenty of motivation
for studying quantum information and entanglement. 
Theoretical studies are also important 
in the context of atom-light interactions inside structured reservoirs \cite{Lambropoulos2000}
such as resonant cavities or photonic band gap materials. The theoretically predicted atom-photon 
bound state could also lead to entanglement and
this can also be linked to another problem: that of atom-laser out-coupling  from 
Bose-Einstein condensates \cite{Nikolopoulos2003,Lazarou2007,Nikolopoulos2008}, where analogous effects
were predicted in the past.  

When quantifying entanglement between an atom and a reservoir of modes, the modes can be treated collectively \cite{CummingsN2008}.
The system is described in terms of two subsystems and one can use existing bipartite entanglement measures. This of course does not permit the study
of entanglement between individual reservoir modes. It is also possible to partition the reservoir and then quantify entanglement between different
parts of the reservoir \cite{Leandro2009}. A different approach is that offered by a recently proposed measure, the density of entanglement \cite{Lazarou2011}.
This measure quantifies entanglement between the atom and different modes in terms of time-dependent distributions.

The problem of entanglement between an atom and a bath of modes, is becoming more interesting when considering 
reservoirs with a spectral gap in their densities of states. For such systems, it is well known that an atom-photon bound state can be formed \cite{Lambropoulos2000,Nikolopoulos1999,Nikolopoulos2000,Bay1997,Bay1998,John1994,Vats1998,Kofman1994}. 
In view of this result, it is reasonable to expect strong quantum correlations and entanglement between the atom and the 
reservoir. 

Motivated by this we consider here a two-level atom coupled to a model reservoir with a single frequency gap in its density of modes.
Exploring the dynamics at different coupling regimes, we are able to show that when a trapping state is formed, permanent correlations are observed. Using the pseudomodes technique \cite{Garraway1997a,Garraway1997b}, and a tripartite entanglement measure, 
the tangle \cite{Coffman2000}, we quantify and study the properties of entanglement. Furthermore, a careful analysis reveals that in the long time
limit and when a trapping state is formed, the reservoir spectrum is not constant in time. This is due to a continuous coupling between the atom and individual
modes, which has zero net energy flow, but induces a permanent effective coupling between the reservoir modes. 
In terms of the pseudomode description, the population trapping arises because of the dark state between the atom and one of the pseudomodes.

This paper is organised as follows. In section \ref{sec:2}, we introduce the model and the pseudomodes method. In section  \ref{sec3}, we discuss the formation
of the trapping state and the reservoir dynamics in the long time limit. In section \ref{sec4}, an analysis of entanglement dynamics in terms of the tangle and 
the density of entanglement is presented. We conclude in section \ref{conclusion}, and in appendix \ref{secA1} a synopsis of the pseudomodes method is provided. 

\section{Model} \label{sec:2}
The system we consider in this work, consists of a two-level atom coupled to a reservoir of harmonic
oscillators with annihilation and creation operators $\hat{a}_\lambda$ and
$\hat{a}_\lambda^\dagger$ respectively. Within the rotating wave approximation the
Hamiltonian reads $(\hbar=1)$
\begin{equation} \label{eq1}
  \begin{split}
    H=&\sum_{\lambda} \omega_\lambda \hat{a}_\lambda^\dagger \hat{a}_\lambda+\omega_0\vert
    1_a\rangle\langle 1_a\vert
    \\ \\
    &+\sum_{\lambda}g_\lambda\left(\hat{a}_\lambda^\dagger
    \vert0_a\rangle\langle1_a\vert+\hat{a}_\lambda\vert1_a\rangle\langle0_a\vert\right),
  \end{split}
\end{equation}
where $g_\lambda$ is the coupling between the mode $\lambda$ and the atomic
transition $\vert1_a\rangle\rightarrow\vert0_a\rangle$. The atomic transition
frequency is $\omega_0$ whereas the $\lambda$-mode frequency is $\omega_\lambda$. 

For the purposes of the analysis that follows, it is very useful to introduce the
reservoir structure function $D(\omega_\lambda)$ which reflects the properties
of the density of modes \cite{Garraway1997a}. This is defined through
\begin{equation} \label{eq2}
  \rho_\lambda(g_\lambda)^2=\frac{\Omega^2_0}{2\pi}D(\omega_\lambda),
\end{equation}
and is normalized such that
\begin{equation} \label{eq3}
  \int_{-\infty}^{\infty}\textrm{d}\omega D(\omega)=2\pi.
\end{equation}
With this normalization a measure of the overall coupling strength is $\Omega_0$ which is given by
\begin{equation} \label{eq4}
  \Omega_0^2=\sum_{\lambda} (g_\lambda)^2.
\end{equation}
In Eq. (\ref{eq2}) $\rho_\lambda$ is the density of modes i.e. the number of modes with frequencies 
in the interval $\omega_\lambda$ to $\omega_\lambda+\textrm{d}\omega_\lambda$.

Previous studies revealed that the formation of an atom-photon bound state is plausible, when an atom is coupled to a reservoir with a gap in its structure function \cite{Lambropoulos2000,Nikolopoulos1999,Nikolopoulos2000,Bay1997,Bay1998,John1994,Vats1998,Kofman1994}. It has also been suggested that the formation of such a bound state is an indication of entanglement between the atom and its environment \cite{Entezar2009,Jiang2011}. In order to explore entanglement dynamics between an atom and a reservoir with a gap at a given frequency $\omega_c$, we utilise the following structure function for the reservoir
\begin{equation}\label{eq5}
D(\omega)=W_1\frac{\Gamma_1}{(\omega-\omega_c)^2+(\Gamma_1/2)^2}-W_2\frac{\Gamma_2}{(\omega-\omega_c)^2+(\Gamma_2/2)^2}.
\end{equation}
This superposition of Lorentzians with the same centre frequency $\omega_c$, widths $\Gamma_j$, amplitudes $\textrm{W}_j$ and opposite signs will result in a gap,  i.e. $\textrm{D}(\omega_c)=0$, if $\Gamma_1 W_2=\Gamma_2 W_1$. Because of the normalisation condition (\ref{eq3}) we also have that $\textrm{W}_1-\textrm{W}_2=1$.

Starting with the atom initially excited and the reservoir in the vacuum state, one has to solve the Schr\"odinger equation 
to obtain the system dynamics for $t>0$. This can be done either with analytical methods, e.g. the Laplace transform 
\cite{John1994,Kofman1994}, or numerical integration \cite{Lambropoulos2000,Nikolopoulos1999,Nikolopoulos2000}.
An alternative approach is that offered by the pseudomodes method \cite{Garraway1997a,Garraway1997b}.
\begin{figure}
\begin{center}
\includegraphics[width=\columnwidth]{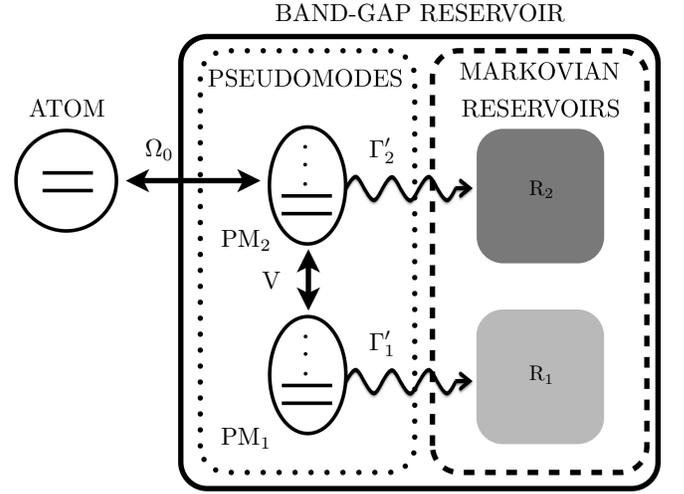}
\caption{Diagrammatic representation of the atom-pseudomodes system. The band-gap reservoir is represented by two interacting pseudomodes PM$_1$ and PM$_2$. The two pseudomodes are also coupled to two independent Markovian reservoirs, that induce the decay of the pseudomodes at rates $\Gamma'_1$ and $\Gamma'_2$ respectively. The atom couples only to one of the two pseudomodes. For a perfect gap i.e. $D(\omega_c)=0$, the decay rate for the first pseudomode $\Gamma'_1$ is zero.}\label{diagram}
\end{center}
\end{figure}

According to this method the reservoir modes are replaced by two degenerate pseudomodes \cite{Garraway1997a,Garraway1997b}, see figure \ref{diagram}. The two pseudomodes are interacting with each other, while one of them is also coupled to the atom. Finally the two pseudomodes decay at rates $\Gamma'_1$ and $\Gamma'_2$ respectively. 

The dynamics of the system are described by a Markovian master equation \cite{Garraway1997a,Garraway1997b}
\begin{equation}\label{eq11}
\begin{split}
\dot{\rho}(t)=&-\imu\left[H_0,\rho(t)\right]-\sum_{j=1}^{2}\frac{\Gamma'_j}{2}\Big(\hat{a}_j^\dagger\hat{a}_j\rho(t) \\ \\
	&-2\hat{a}_j\rho(t)\hat{a}^\dagger_j+\rho(t)\hat{a}^\dagger_j\hat{a}_j\Big),
\end{split}
\end{equation}
with the Hamiltonian
\begin{equation}\label{eq12}
\begin{split}
H_0=&\omega_0\vert1_a\rangle\langle1_a\vert+\omega_c\left(\hat{a}^\dagger_1\hat{a}_1+\hat{a}^\dagger_2\hat{a}_2\right) \\ \\
&+\Omega_0\left(\hat{a}^\dagger_2\vert0_a\rangle\langle1_a\vert+\hat{a}_2\vert1_a\rangle\langle0_1\vert\right) \\ \\
&+V\left(\hat{a}^\dagger_1\hat{a}_2+\hat{a}_1\hat{a}^\dagger_2\right),
\end{split}
\end{equation}
where $\hat{a}_1$ $(\hat{a}^\dagger_1)$ and $\hat{a}_2$ $(\hat{a}^\dagger_2)$ are the annihilation (creation) operators for the
two pseudomodes respectively. The vacuum and excited states for the atom are $\vert0_a\rangle$ and $\vert1_a\rangle$.
The coupling $\Omega_0$ is given by Eq. (\ref{eq4}), and $V=\sqrt{W_1W_2}(\Gamma_1-\Gamma_2)/2$. The two decay rates are
$\Gamma'_1=W_1\Gamma_2-W_2\Gamma_1$ and $\Gamma'_2=W_1\Gamma_1-W_2\Gamma_2$. For a perfect gap $D(\omega_c)=0$,
the decay rate for the first pseudomode is $\Gamma'_1=0$, $\Gamma'_2=(\Gamma_1+\Gamma_2)$ and $V=\sqrt{\Gamma_1\Gamma_2}/2$.

The solution for the master equation (\ref{eq11}) reads 
\begin{equation}\label{eq13}
\rho(t)=\Pi_j(t)\vert0_a0_10_2\rangle\langle0_a0_10_2\vert+\vert\tilde{\psi}(t)\rangle\langle\tilde{\psi}(t)\vert, 
\end{equation}
where 
\begin{equation}\label{eq14}
\vert\tilde{\psi}(t)\rangle=c_a(t)\vert1_a0_10_2\rangle+a_1(t)\vert0_a1_10_2\rangle+a_2(t)\vert0_a0_11_2\rangle.
\end{equation}
The vacuum state $\vert0_a0_10_2\rangle$ population is $\Pi_j(t)$, and the probability amplitudes for the atom, and the first and second
pseudomodes are $c_a(t)$, $a_1(t)$ and $a_2(t)$ respectively. A synopsis of the pseudomode method is provided in the appendix
\ref{secA1}, along with the expressions for the probability amplitudes $c_a(t)$, $a_1(t)$, $a_2(t)$ and the population $\Pi_j(t)$.
\section{Population trapping and non-stationary environment}\label{sec3}
\begin{figure}
\begin{center}
  \subfigure{\includegraphics[width=0.8\columnwidth]{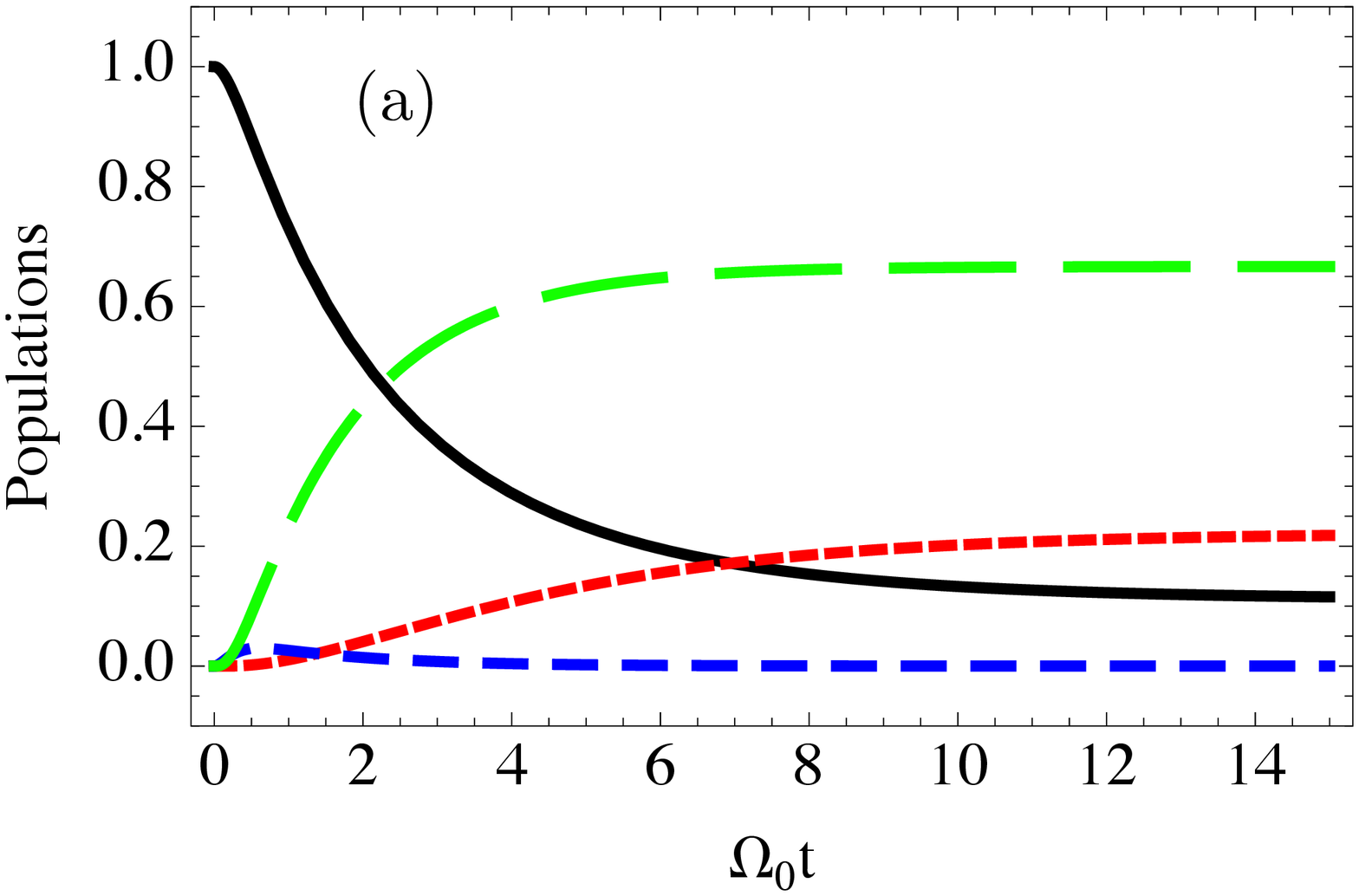}\label{fig1a}} 
  
  \subfigure{\includegraphics[width=0.8\columnwidth]{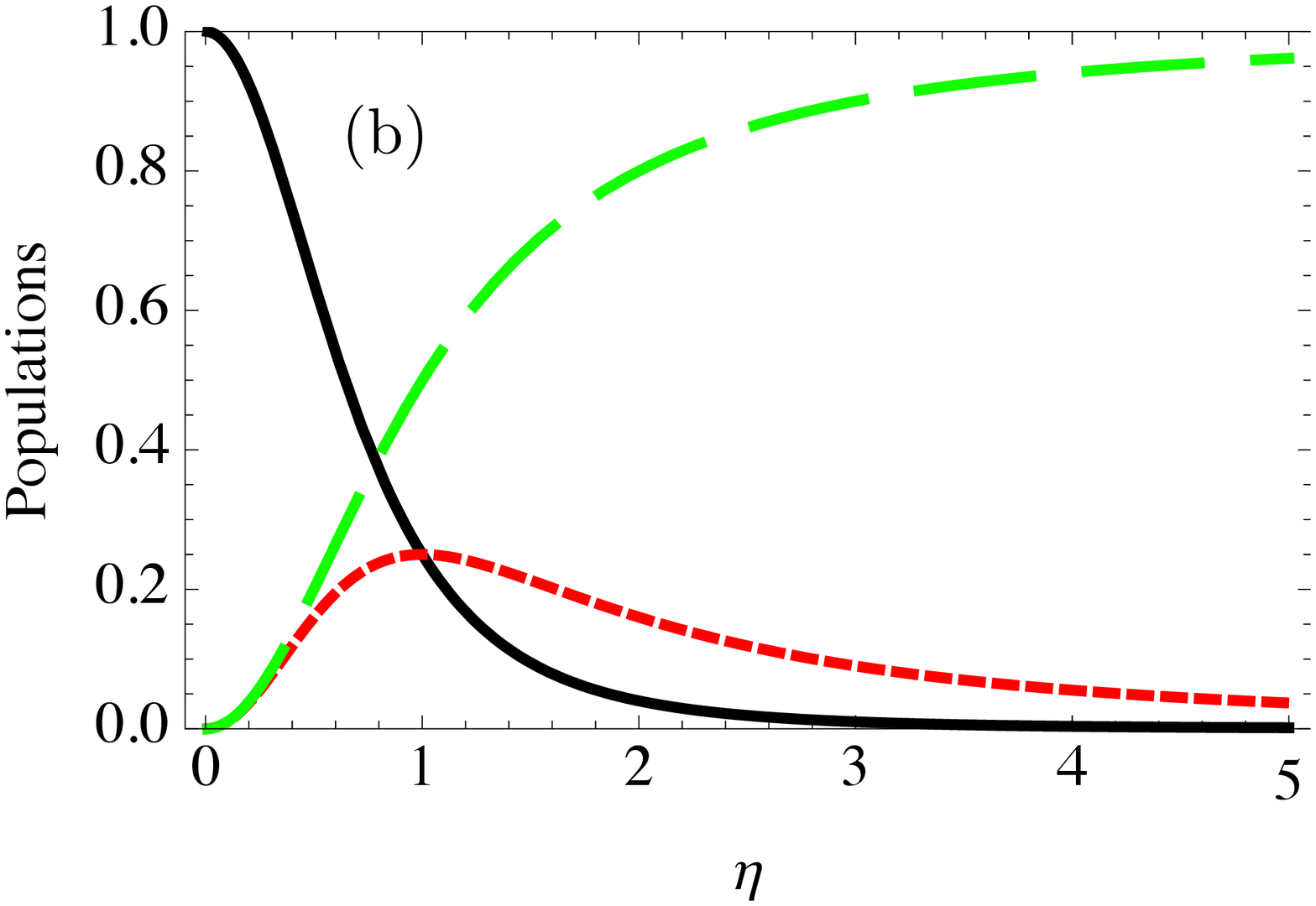}\label{fig1b}} 
    \caption{(Color online) (a) The populations $\vert c_a(t)\vert^2$ (black), $\vert a_1(t)\vert^2$ (red short-dashed), $\vert a_2(t)\vert^2$ (blue dashed) and $\Pi_j(t)$ (green long-dashed),  	for $\Gamma_1=10\Omega_0$, $\Gamma_2=0.2\Omega_0$ and $W_1=50W_2$. (b) The final populations $\vert c_a(\infty)\vert^2$ (black), $\vert a_1(\infty)\vert^2$ (red 	short-dashed) and $\Pi_j(\infty)$ (green-dashed) as functions of the dimensionless parameter $\eta=2\Omega_0/\sqrt{\Gamma_1\Gamma_2}$. }
    \label{fig1}
\end{center}
\end{figure}

For a resonant system $\omega_c=\omega_0$, and when the perfect gap condition $\Gamma_1 W_2=\Gamma_2 W_1$ is met, we have that $D(\omega_c)=0$ and in the long time limit a trapping state is formed \cite{Garraway1997b}. Upon solving the equations for $c_a(t)$, $a_1(t)$ and $a_2(t)$, see appendix \ref{secA1}, and taking their limits for $t\rightarrow\infty$ we have that
\begin{equation}\label{eq16}
c_a(\infty)=(1+\eta^2)^{-1},
\end{equation}
and 
\begin{equation}\label{eq17}
a_1(\infty)=\eta(1+\eta^2)^{-1},
\end{equation}
where $\eta=2\Omega_0/\sqrt{\Gamma_1\Gamma_2}$. The probability amplitude for the second pseudomode is $a_2(\infty)=0$ and the population of the vacuum state is 
\begin{equation}\label{eq18}
\Pi_j(\infty)=\eta^2(1+\eta^2)^{-1}.
\end{equation}
A plot of $|c_a(t)|^2$, $|a_1(t)|^2$, $|a_2(t)|^2$ and $\Pi_j(t)$ for $\Gamma_1=10\Omega_0$ and $\Gamma_2=0.2\Omega_0$ is shown in figure \ref{fig1a}.

From the above three equations it is evident that in the long time limit, and in the weak coupling regime $\eta\ll1$, a fraction of the population will remain trapped in the excited atomic state and the the second pseudomode, see figure \ref{fig1b}. The remaining population is irreversibly lost to the reservoir (or more
  precisely, to the Markovian part of the reservoir \cite{Mazzola2009}).  From figures \ref{fig1b} and Eqs. (\ref{eq16})-(\ref{eq18}) we can see that population trapping, i.e. $\vert c_a(\infty)\vert^2$, is significant for
$\eta\leq1$. The population lost to the reservoir, i.e. the sum of the populations for the pseudomode 1 $\vert a_1(\infty)\vert^2$ and the vacuum state $\Pi_j(t)$, remains low for $\eta\ll1$, see Fig. \ref{fig1b}. As we move to the strong coupling regime $\eta\ge1$ losses increase, and eventually for $\eta\gg1$ all the population is transferred to the reservoir. 

An interesting feature of the trapping state, is that in the long time limit the reservoir modes do not reach a steady state.  This can be evidenced in the reservoir spectrum for $t\rightarrow\infty$. Using the definition for the reservoir spectrum  \cite{Linington2006}
\begin{equation} \label{eq19}
  S(\omega_\lambda,t)=\rho_\lambda\vert c_{\lambda}(t)\vert^2,
\end{equation}
and Eq. (\ref{eqA17}) for $t(\Gamma_1+\Gamma_2)\gg1$ we get the following expression for $S(\omega_\lambda,t)$
\begin{equation}\label{eq20}
\begin{split}
S(\omega_{\lambda},t)=\frac{8\Omega^2_0D(\omega_{\lambda})}{\pi(4\Gamma^2+\Omega^2)^2}\Bigg\vert &\frac{\Gamma_1\Gamma_2}{2\delta_\lambda}e^{\imu\delta_\lambda t/2}\sin\left(\frac{\delta_\lambda t}{2}\right) \\  &+\frac{4\Omega^2_0\left(2\Gamma-\imu\delta_\lambda\right)}{4(\Gamma-\imu\delta_\lambda)^2+\Omega^2}\Bigg\vert^2,
\end{split}
\end{equation}
where the width $\Gamma$ and the Rabi frequency $\Omega$ are given in Eqs. (\ref{eqA14}) and (\ref{eqA15}), and $\delta_\lambda=\omega_\lambda-\omega_c$.

\begin{figure*}
  \begin{center}
    \subfigure{\includegraphics[width=0.8\columnwidth]{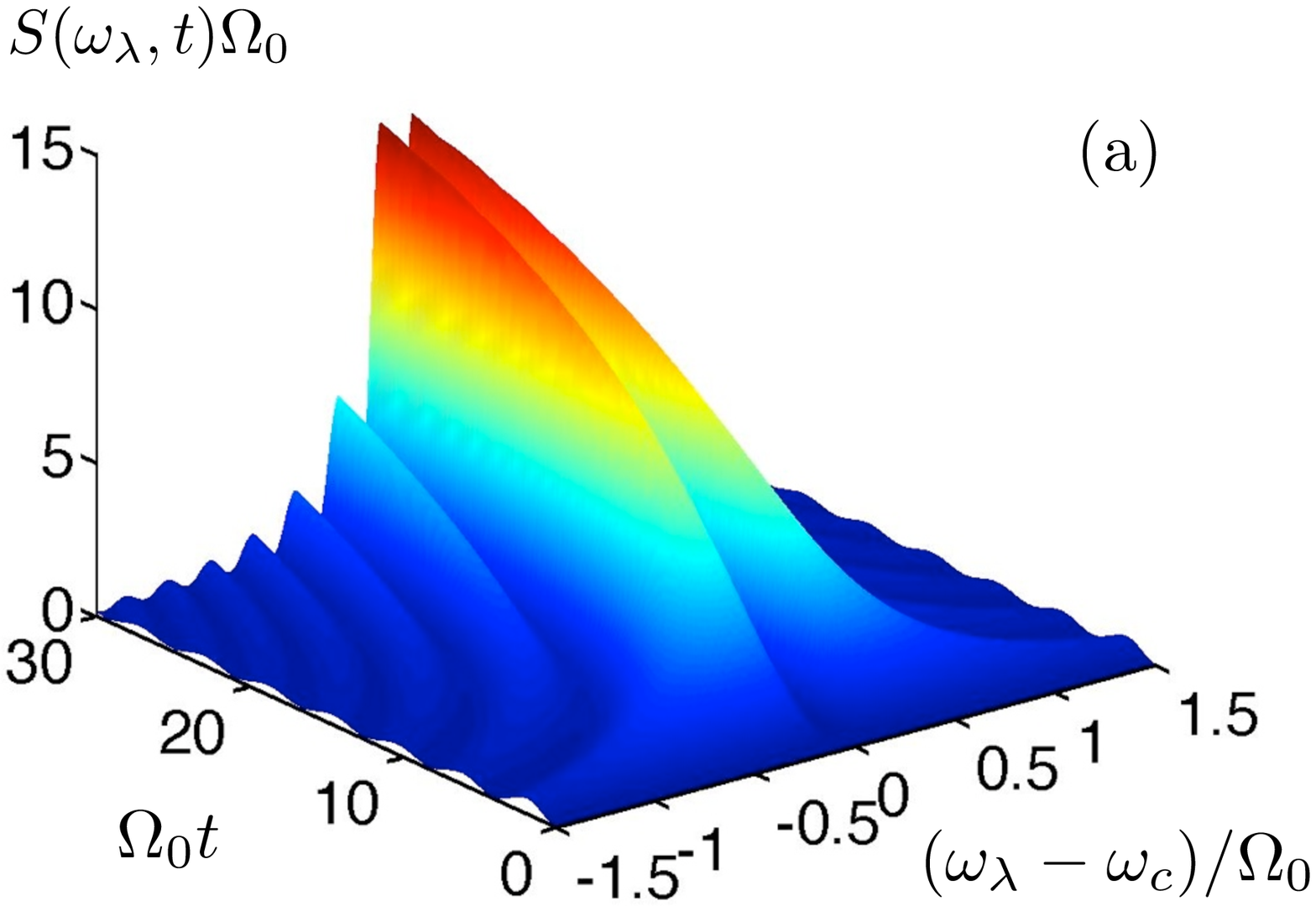}\label{fig2a}} \qquad\qquad
    \subfigure{\includegraphics[width=0.8\columnwidth]{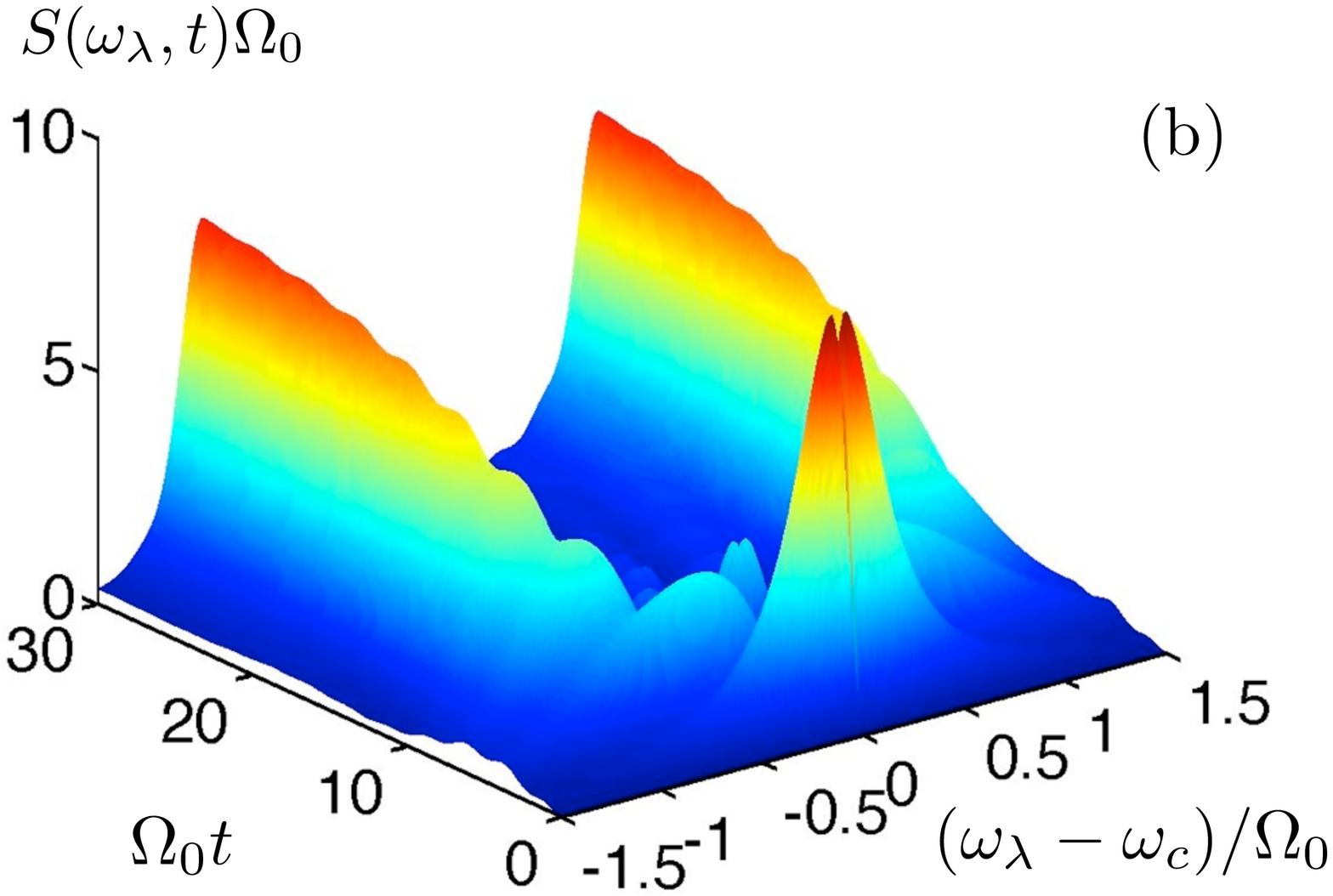}\label{fig2b}}\\
    \subfigure{\includegraphics[width=0.8\columnwidth]{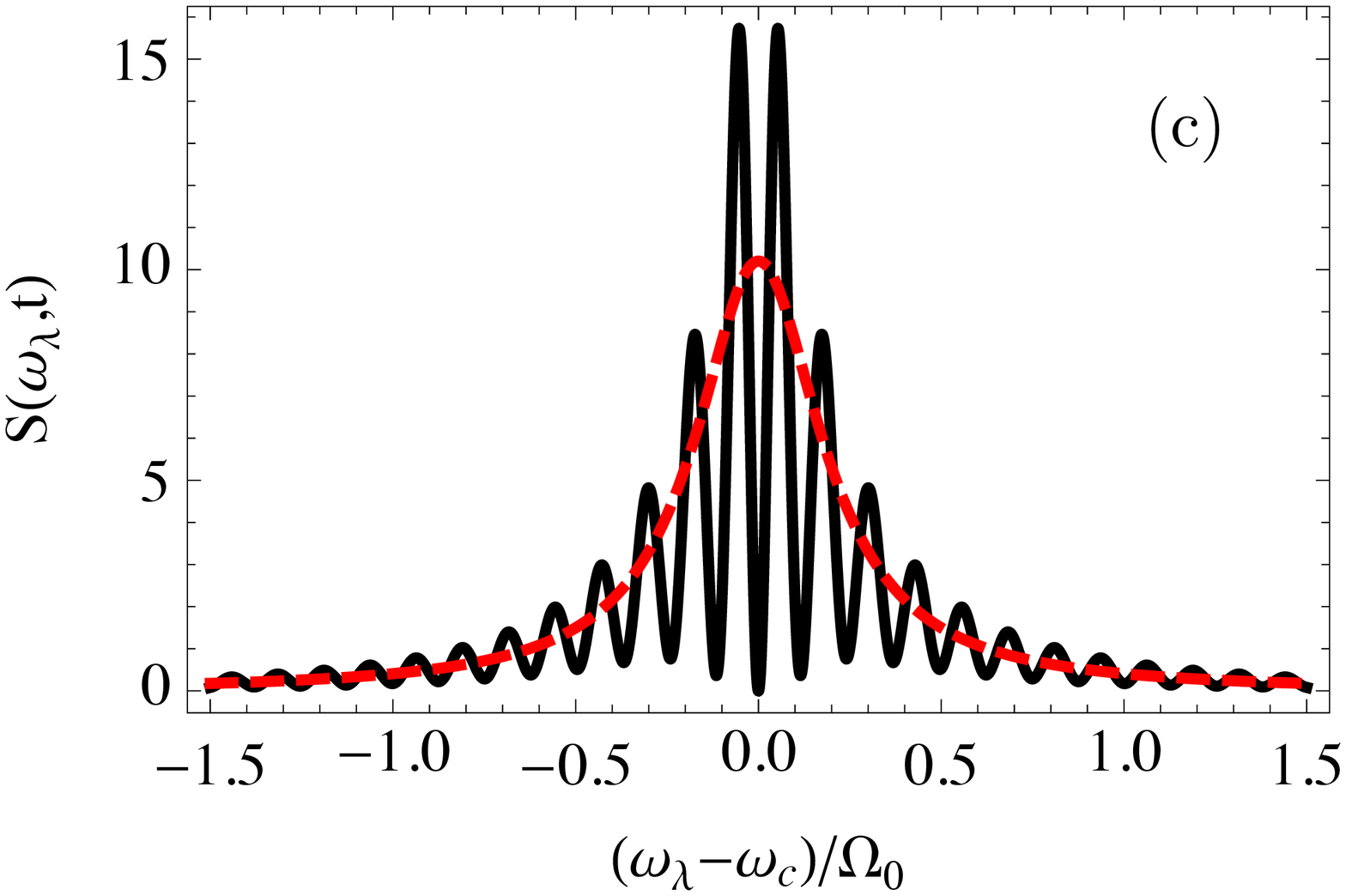}\label{fig2c}} \qquad\qquad
    \subfigure{\includegraphics[width=0.8\columnwidth]{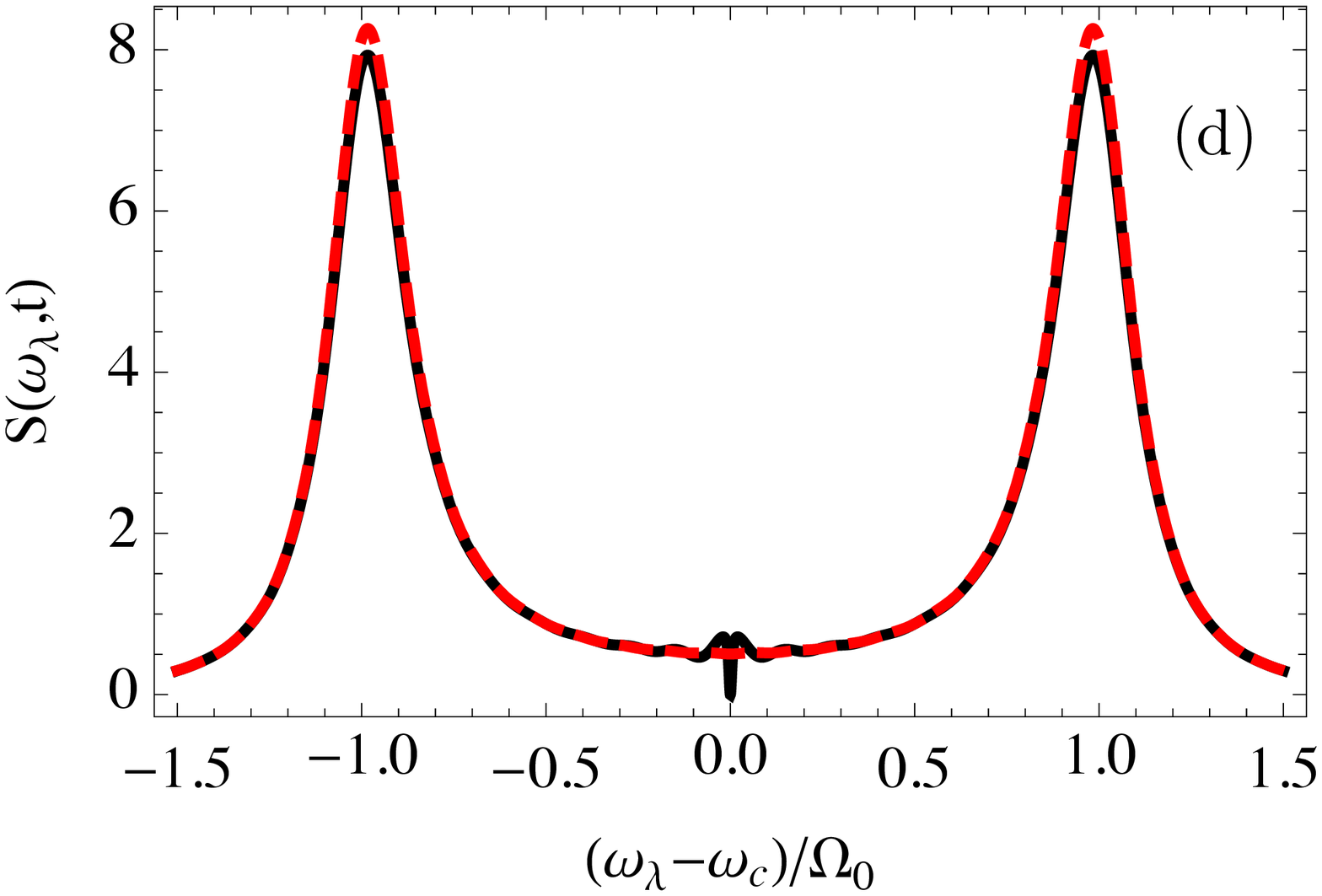}\label{fig2d}}
    \caption{(Color online) The reservoir spectrum $S(\omega_\lambda,t)$ as a function of time, (a) for $\Gamma_1=10\Omega_0$, $\Gamma_2=0.2\Omega_0$ and $W_1=50W_2$, and (b)  for $\Gamma_1=0.5\Omega_0$, $\Gamma_2=0.01\Omega_0$ and $W_1=50W_2$. Figures (c) and (d) are snapshots for the reservoir spectrum for $\Omega_0t=50$ and for the parameters of figures (a) and (b) respectively. The red dashed lines in figures (c) and (d) is the spectrum for a reservoir with a Lorentzian structure function with $\Gamma_2=W_2=0$, $W_1=1$ and $\Gamma_1=10\Omega_0$ and $\Gamma_1=0.5\Omega_0$ respectively.}
    \label{fig2}
  \end{center}
\end{figure*}

Thus in the long time limit, although the total excitation in the reservoir is constant, the modes remain coupled to each other. As a result of this the population distribution between the modes changes, see Fig.~\ref{fig2}.
The oscillatory exchange of population between the modes is more pronounced in the weak coupling regime $\Omega_0\ll\sqrt{\Gamma_1\Gamma_2}$, Fig.~\ref{fig2a}, and is negligible for the strong coupling regime, Fig.~\ref{fig2b}.
Snapshots of the reservoir spectrum for times  $t(\Gamma_1+\Gamma_2)\gg1$ are shown in Figs.~\ref{fig2c} and \ref{fig2d}.
It is also interesting to note that in the weak coupling regime displaying the population trapping, the frequency gap imposes strong oscillations in the mode populations compared to the single Lorentzian structure function case\cite{Lazarou2011}, see Fig.~\ref{fig2c}. In contrast with strong coupling and no population trapping, there is a strong resemblance in the mode populations between the gap and single Lorentzian cases (Fig.~\ref{fig2d}).

\begin{figure}
  \begin{center}
    \subfigure{\includegraphics[width=0.8\columnwidth]{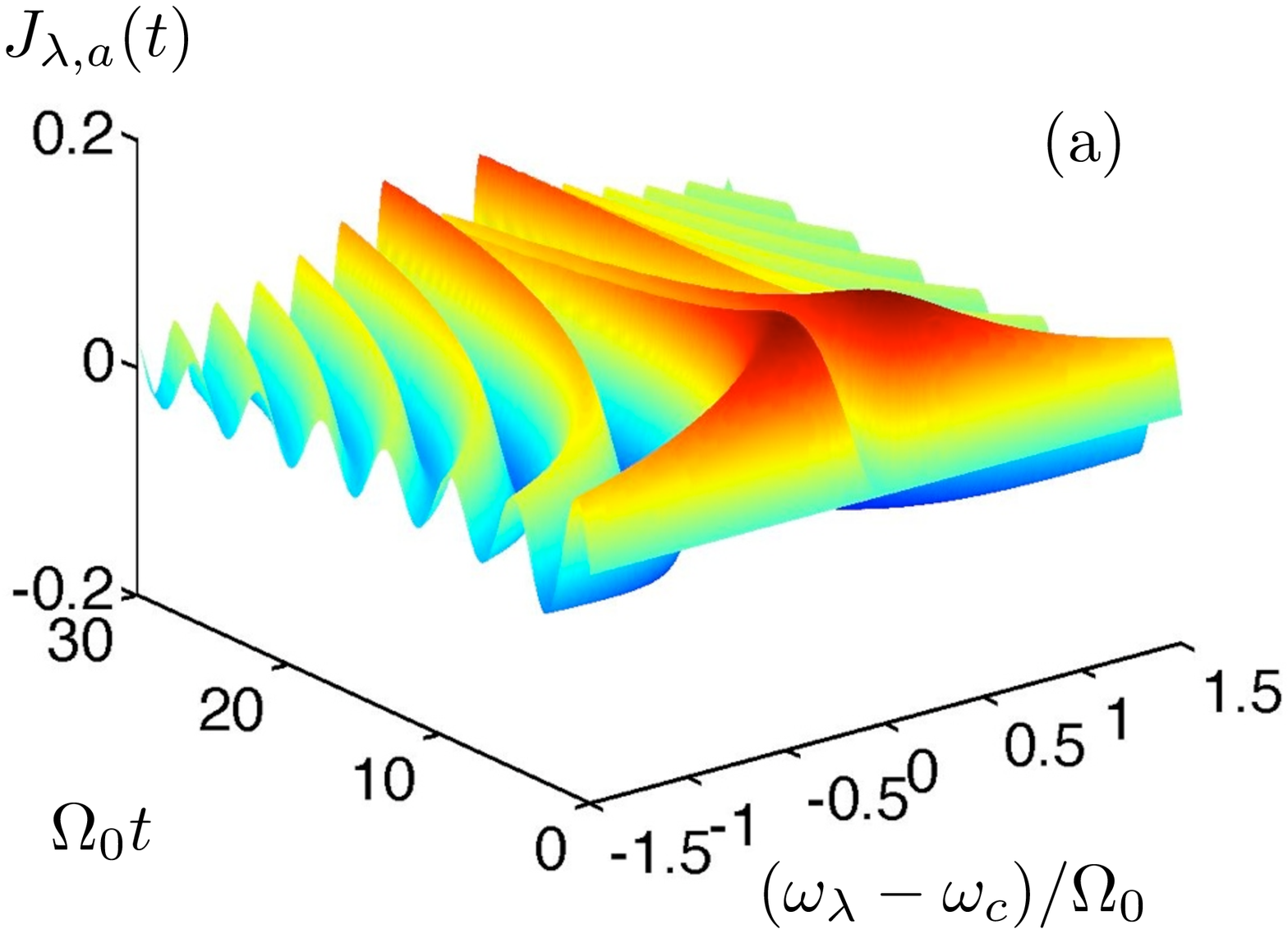}\label{fig3a}}

    \subfigure{\includegraphics[width=0.8\columnwidth]{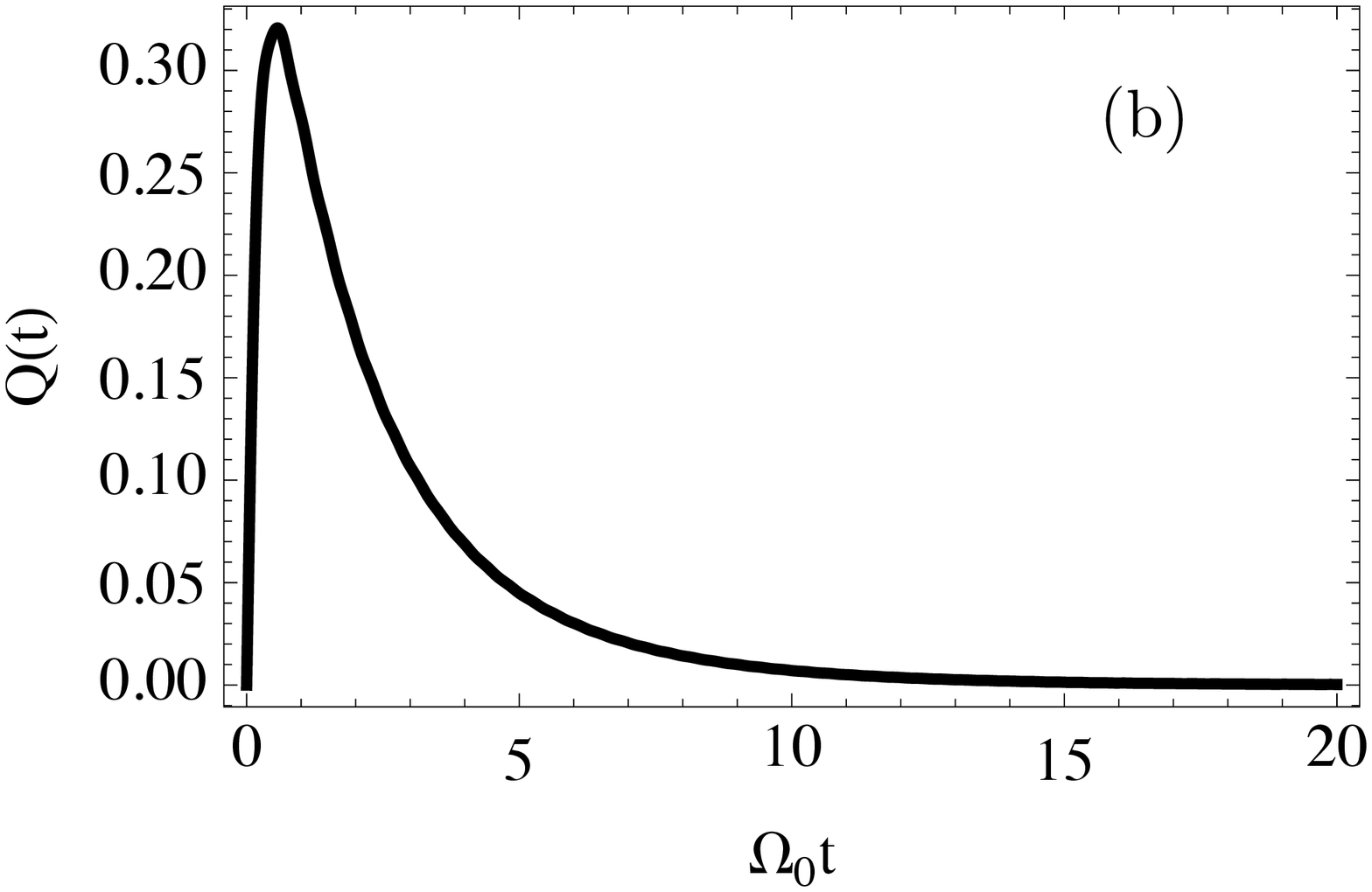}\label{fig3b}}
    \caption{(Color online) The probability current $J_{\lambda,a}(t)$ Eq. (\ref{eq20}) (a), for $\Gamma_1=10\Omega_0$, $\Gamma_2=0.2\Omega_0$ and $W_1=50W_2$, and the total probability current $Q(t)$ Eq. (\ref{eq21}) for the same parameters (b).}
    \label{fig3}
  \end{center}
\end{figure}

In order to explore further these features, we plot in Fig.~\ref{fig3a} the probability current between the atom and the $\lambda$ mode \cite{Gambetta2004,Luoma2011}
\begin{equation}\label{eq21}
J_{\lambda,a}(t)=2\textrm{Im}\big\{\rho_\lambda g_\lambda\tilde{c}_{\lambda}^*(t)\tilde{c}_a(t)e^{\imu\delta_\lambda t}\big\},
\end{equation}
and in Fig.~\ref{fig3b}
the net probability current
\begin{equation}\label{eq22}
Q(t)=\int_{-\infty}^{\infty}\textrm{d}\omega_\lambda J_{\lambda,a}(t).
\end{equation}
The long time limit, i.e. when $t\rightarrow\infty$, is of particular interest.  We see that while the net probability current
Fig. \ref{fig3b} approaches zero, in the long time limit, the individual reservoir frequency components seen in Fig. \ref{fig3a}
do not decay, but continue oscillating. This does not happen in the case of a Lorentzian reservoir coupling and appears to 
be a feature of population trapping in a photonic band-gap structure. Subsequently, there exists an effective, atom-mediated, coupling between the modes 
even though the atom has reached a steady state.

For reservoirs with a single Lorentzian structure function, the pseudomodes method has provided an intuitive insight into memory effects \cite{Mazzola2009}. When an atom is coupled to such an environment, slowly decaying oscillations between the atom and the reservoir are observed in the strong coupling limit \cite{Mazzola2009,Lazarou2011}. These correlation effects, are attributed to a memory part of the reservoir that is represented by a single pseudomode. The rest of the reservoir acts as a Markovian environment that induces a slow exponential decay for the memory part. 

In the current system with a frequency gap in the environment, we have 
two pseudomodes which store information about the state of the atom.
The first one $a_1(t)$, i.e. the one that forms the trapping state with the atom, is responsible for the permanent storage of information. The second one $a_2(t)$ is responsible for short term storage, but eventually gets depleted due to its coupling to the rest of the reservoir. Though the atom is directly coupled to the first pseudomode only, the interaction between the pseudomodes gives rise to the trapping of population by forming a dark state for the atom - pseudomode one subsystem.  
 
In general, the population trapping signifies the formation of an atom-photon bound state. In view of the strong permanent correlation effects that dictate the formation of such a state, it is reasonable to expect entanglement to be also present. In the following section we use both the pseudomodes method, and the recently proposed density of entanglement \cite{Lazarou2011}, to explore entanglement between the atom and the reservoir defined by Eq.~(\ref{eq5}).

\section{Entanglement dynamics}\label{sec4}
Identifying and measuring entanglement in multi-partite systems presents various complications. Apart from the case of a two-qubit system, where entanglement can be identified both for a pure and a mixed state \cite{Hill1997,Wootters1998}, multi-qubit entanglement is an open problem and to date several measures of entanglement have been proposed 
\cite{Amico2008,Horodecki2009,Mintert2005a,Barnum2004,Meyer2002,Aktharshenas2005,Lazarou2011,Coffman2000}. For the analysis that follows, we will be using two different measures \cite{Coffman2000,Lazarou2011}.  

The first one, called tangle \cite{Coffman2000}, is a measure of genuine tripartite entanglement between three qubits. This will be used to explore entanglement dynamics in the pseudomodes framework. The second one is the recently proposed density of entanglement \cite{Lazarou2011}. This measure is appropriate for studying entanglement between an atom and the continuum of the reservoir modes. It provides valuable information regarding entanglement distribution between the atom and the modes and between individual modes.
\subsection{Tangle} 
We start our analysis from Eqs. (\ref{eq13}) and (\ref{eq14}), i.e. the density matrix for the atom-pseudomodes system. For this mixed state, the two pseudomodes can be collectively described in terms of a single qubit. The two states for this collective qubit are 
\begin{equation}\label{eq23}
\vert0_{ps}\rangle=\vert0_10_2\rangle,
\end{equation}
and
\begin{equation}\label{eq24}
\vert1_{ps}\rangle=\frac{1}{\sqrt{\vert a_1(t)\vert^2+\vert a_2(t)\vert^2}}\left(a_1(t)\vert1_10_2\rangle+a_2(t)\vert0_11_2\rangle\right).
\end{equation}
Using these expressions, the state $\vert\tilde{\psi}(t)\rangle$ reads
\begin{equation}\label{eq25}
\vert\tilde{\psi}(t)\rangle=c_a(t)\vert1_a0_{ps}\rangle+\sqrt{\vert a_1(t)\vert^2+\vert a_2(t)\vert^2}\vert0_{a}1_{ps}\rangle,
\end{equation}
and the density matrix $\rho(t)$ becomes
\begin{equation}\label{eq26}
\rho(t)=\Pi_{j}(t)\vert0_a0_{ps}\rangle\langle0_a0_{ps}\vert+\vert\tilde{\psi}(t)\rangle\langle\tilde{\psi}(t)\vert.
\end{equation}
We should note here, that the states $\vert0_{ps}\rangle$ and $\vert1_{ps}\rangle$ are both eigenstates with zero eigenvalues for the reduced density matrix $\rho_{12}(t)=\textrm{tr}_a\{\rho\}$ for the two pseudomodes, where the tracing is over the atomic states $\vert 0_a\rangle$ and $\vert 1_a\rangle$.

Entanglement for this ``two-qubit" mixed state can be quantified in terms of the concurrence \cite{Hill1997,Wootters1998}. This can be associated with tangle, a measure of tripartite entanglement for a system of three qubits A, B, and C \cite{Coffman2000}. The tangle $\tau_{ABC}$ expressed in terms of pairwise concurrences
reads
\begin{equation}\label{eq27}
\tau_{ABC}=C^2_{A(BC)}-C^2_{AB}-C^2_{AC}
\end{equation}
where $C_{AB}$ and $C_{AC}$ are the pairwise concurrences for the qubit A with B and C respectively, whereas $C_{A(BC)}$ is the concurrence for 
qubit A and a qubit (BC) that collectively describes qubits B and C. 

The above equation can also be written as an inequality i.e.
\begin{equation}\label{eq28}
C^2_{A(BC)}\ge C^2_{AB}+C^2_{AC}.
\end{equation}
The meaning of these two equations is that entanglement between the qubit A and the other two qubits, B and C, is manifested through direct entanglement
with each qubit, thus the two concurrences $C_{AB}$ and $C_{AC}$, and through a three-way (tripartite) entanglement i.e. $\tau_{ABC}$.

From the three qubit density matrix Eq. (\ref{eq13}), we derive the reduced density matrices for the atom with each individual pseudomode i.e. 
$\rho_{a,1}$ and $\rho_{a,2}$. Using the concurrence for a two-qubit system \cite{Hill1997,Wootters1998} we obtain the following two expressions for the concurrence for the atom with each pseudomode 
\begin{equation}\label{eq29}
C^2_{a,1}(t)=4\vert c_a(t)\vert^2\vert a_1(t)\vert^2,
\end{equation}
and
\begin{equation}\label{eq30}
C^2_{a,2}(t)=4\vert c_a(t)\vert^2\vert a_2(t)\vert^2.
\end{equation}
The final step is to calculate the concurrence for the density matrix (\ref{eq26}). 

This is the concurrence for the atom and the qubit that collectively describes the two pseudomodes.
The calculation is simple and the concurrence $C^2_{a,(12)}(t)$ is
\begin{equation}\label{eq31}
C^2_{a,(12)}(t)=C^2_{a,1}(t)+C^2_{a,2}(t),
\end{equation}
i.e. the tangle for the atom and the two pseudomodes is zero.
From the definition of the tangle, Eqs. (\ref{eq27}) and (\ref{eq28}), and the above result, we conclude that entanglement between the atom and the pseudomodes is manifested only through two-way entanglement channels. A three way entanglement is completely absent. 
\begin{figure}
  \begin{center}
    \subfigure{\includegraphics[width=0.8\columnwidth]{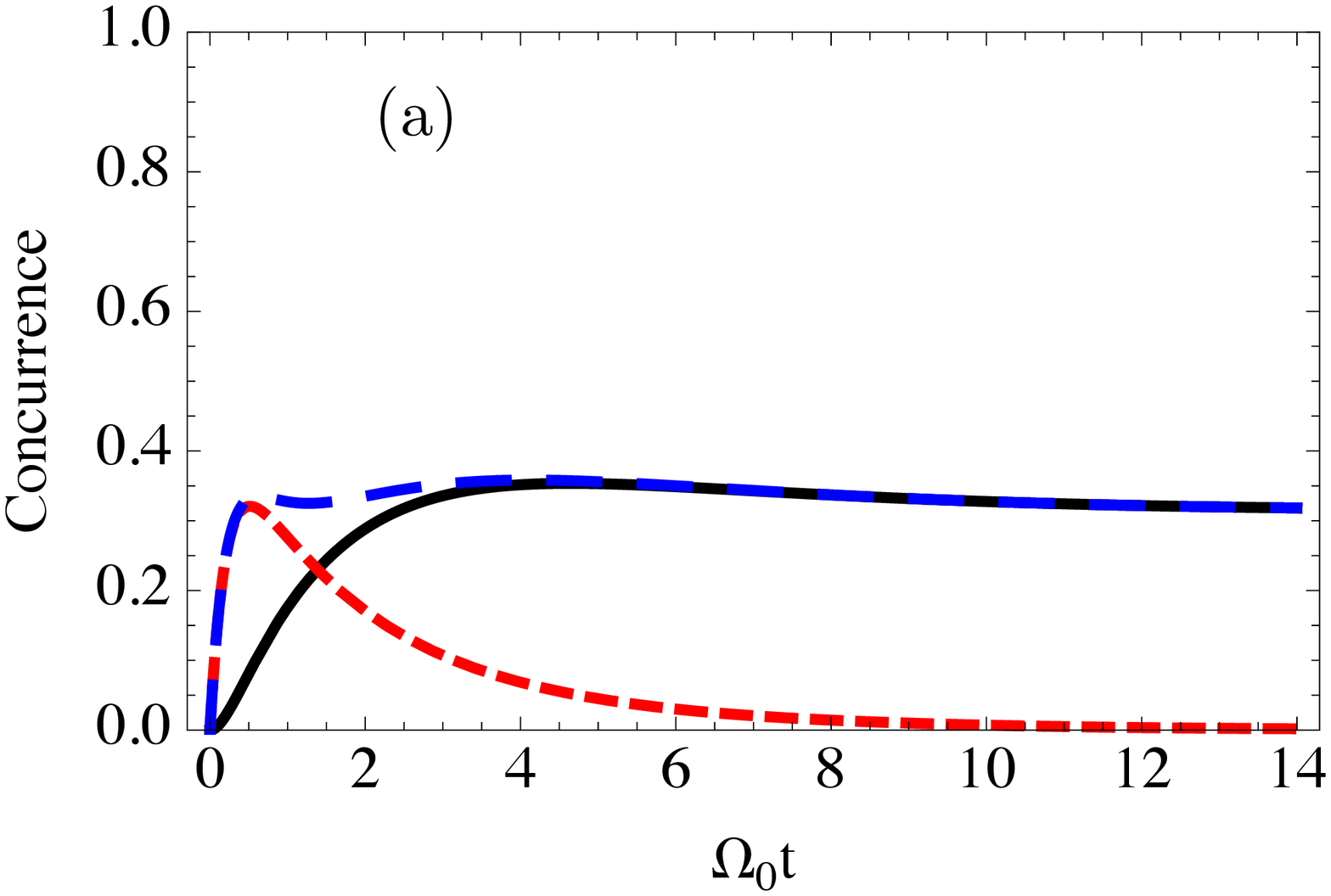}\label{fig4a}}
    
    \subfigure{\includegraphics[width=0.8\columnwidth]{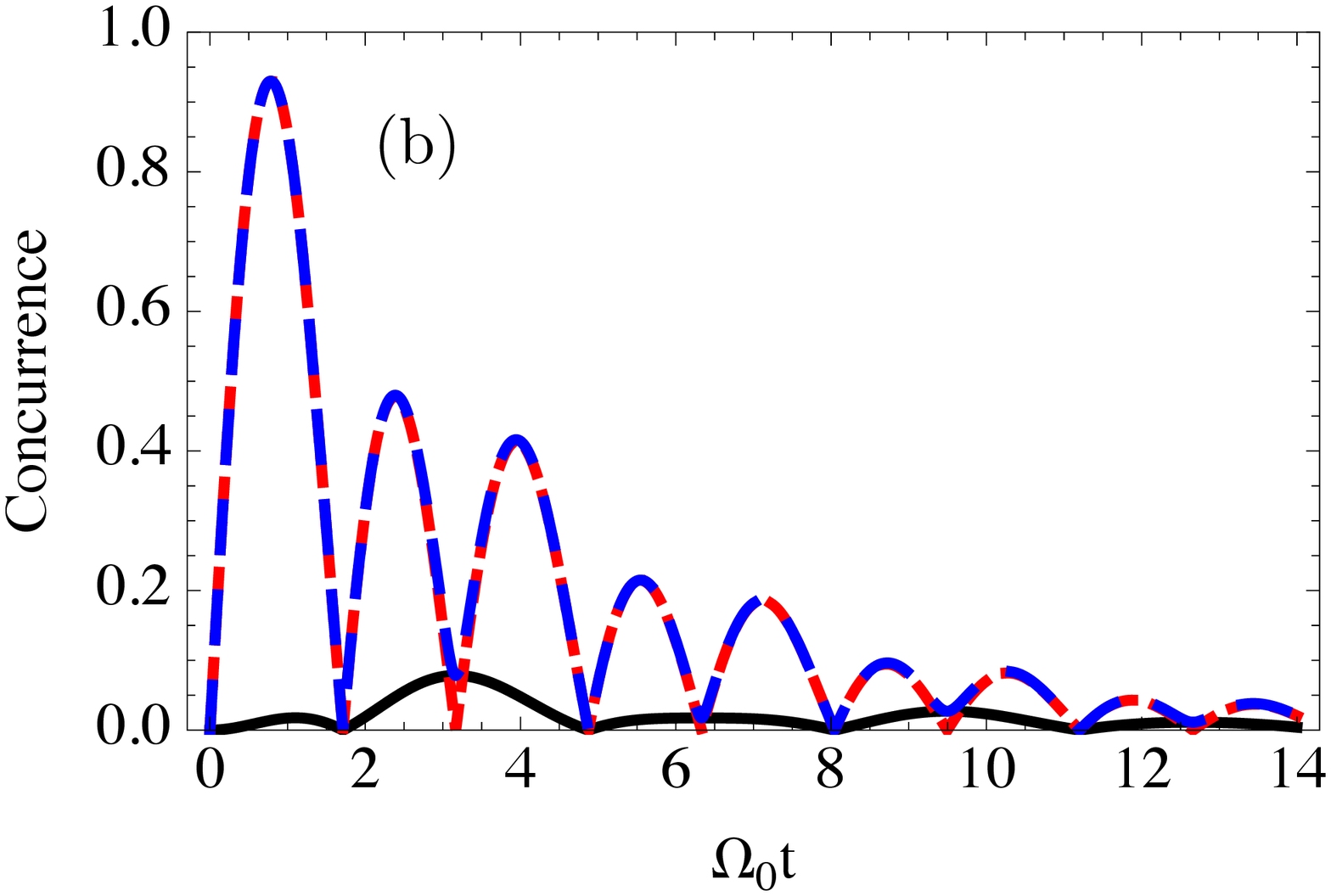}\label{fig4b}}
    \caption{(Color online) The concurrences $C_{a,1}(t)$ (black solid) $C_{a,2}(t)$ (red short-dashed) and $C_{a,(12)}(t)$ (blue dashed),  for $\Gamma_1=10\Omega_0$, $\Gamma_2=0.2\Omega_0$ (a), and for $\Gamma_1=0.5\Omega_0$, $\Gamma_2=0.01\Omega_0$ (b). For both figures $W_1=50W_2$.}
    \label{fig4}
  \end{center}
\end{figure}

In figures \ref{fig4a} and \ref{fig4b}, we plot the concurrences as functions of time, for the weak and strong coupling regimes respectively. In the weak coupling regime, where a trapping state is formed, entanglement at a very early stage builds up only between the atom and pseudomode two which is responsible for the short term storage of information.
 After reaching a peak, it starts decaying where at the same time entanglement between the atom and the pseudomode one, which is responsible for the long-term storage of information, slowly builds up and reaches a steady state. Thus the trapping state is also an entangled state between the atom and the reservoir.

For the strong coupling regime, pseudomode one makes a negligible contribution in the entanglement dynamics. Pseudomode two has a strong contribution for short times, where the concurrence $C_{a,2}(t)$ quickly increases, and then follows a slowly decaying oscillation pattern. These oscillations are the signature of a Rabi splitting observed in the strong coupling regime \cite{Lazarou2011}, see also the reservoir spectrum in figure \ref{fig2b}.
\subsection{Density of entanglement}\label{sec41}
In order to gain further insight into entanglement dynamics, we need to consider entanglement between the atom and each of the reservoir modes.
For quantifying the distribution of entanglement between the atom and the individual reservoir modes, and among the reservoir modes, we use the density of entanglement \cite{Lazarou2011}. The density of entanglement between the atom and modes with frequencies in an interval $\omega_\lambda$ to $\omega_\lambda+\textrm{d}\omega_\lambda$ is 
\begin{equation}\label{eq32}
\mathcal{E}_A(\omega_\lambda,t)=4\vert c_a(t)\vert^2S(\omega_\lambda,t),
\end{equation}
and the density of entanglement among the reservoir modes reads
\begin{equation}\label{eq33}
\mathcal{E}_R(\omega_\lambda,\omega_\mu,t)=2S(\omega_\lambda,t)S(\omega_\mu,t),
\end{equation}
where $S(\omega_\lambda,t)$ is the reservoir spectrum.

In terms of these two distributions the total entanglement or concurrence $C^2(t)$ for the atom and the
reservoir modes is defined as the sum of an atom-modes contribution
\begin{equation}\label {eq34}
C^2_A(t)=\int_{-\infty}^{\infty}\textrm{d}\omega_\lambda\mathcal{E}_A(\omega_\lambda,t),
\end{equation}
and a reservoir contribution 
\begin{equation}\label{eq35}
C^2_R(t)=\int_{-\infty}^{\infty}\textrm{d}\omega_\lambda\int_{-\infty}^{\infty}\textrm{d}\omega_{\mu}\mathcal{E}_{R}(\omega_\lambda,\omega_\mu,t).
\end{equation}
The total entanglement reads
\begin{equation}\label{eq36}
C^2(t)=C^2_A(t)+C^2_R(t).
\end{equation}
\begin{figure*}
  \begin{center}
    \subfigure{\includegraphics[width=0.8\columnwidth]{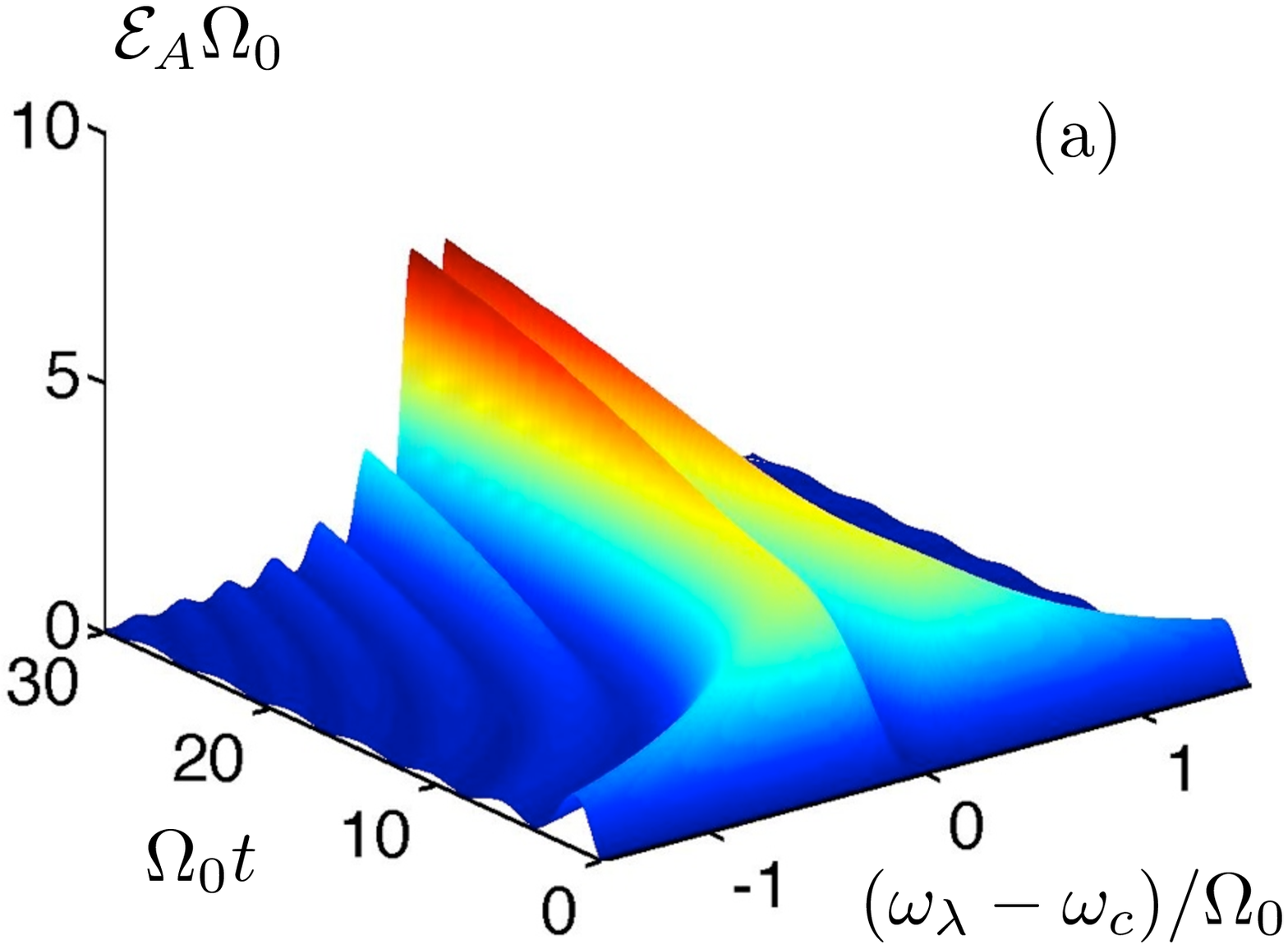}\label{fig5a}}\qquad\qquad
    \subfigure{\includegraphics[width=0.8\columnwidth]{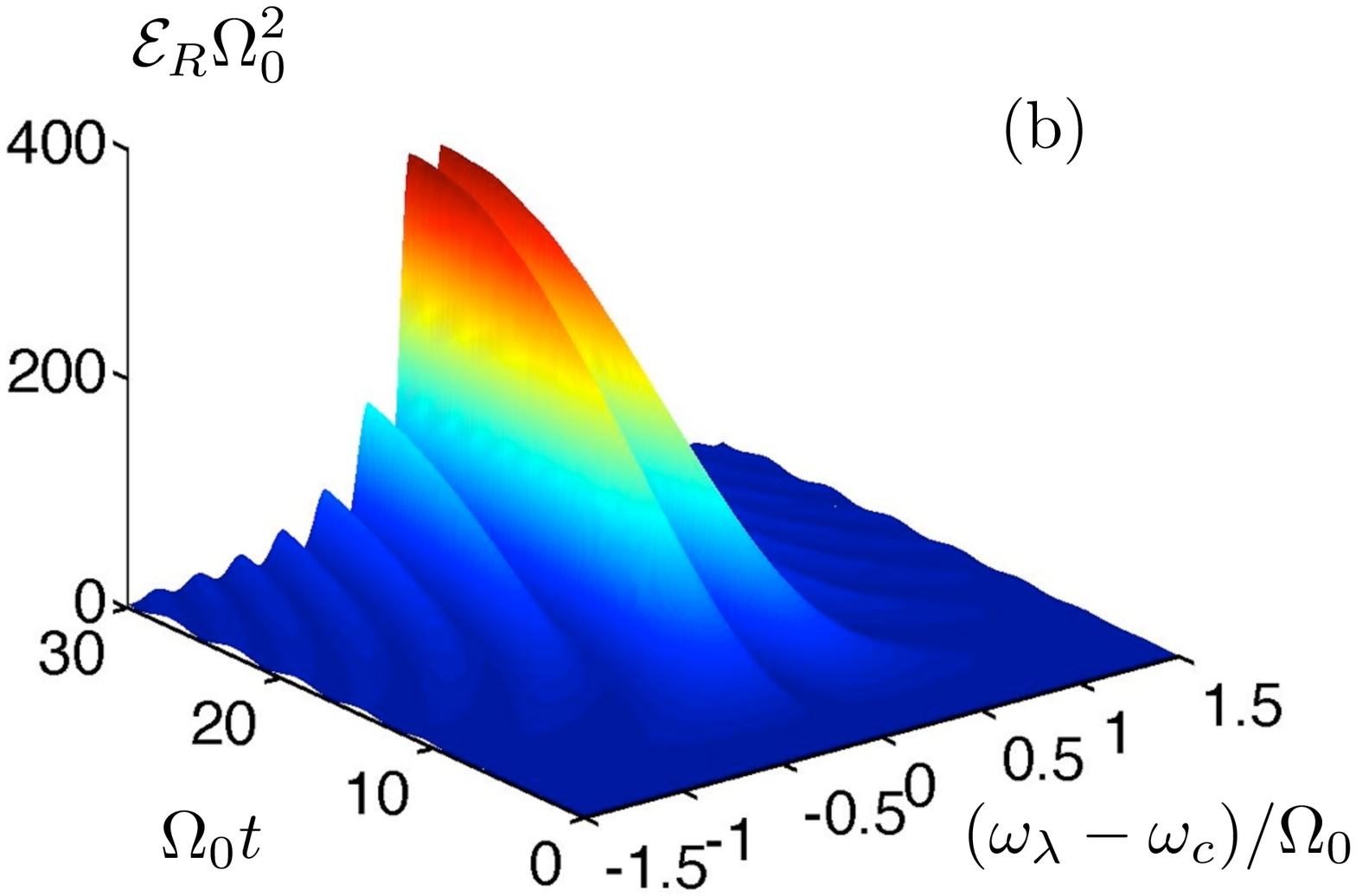}\label{fig5b}}\\
    \subfigure{\includegraphics[width=0.8\columnwidth]{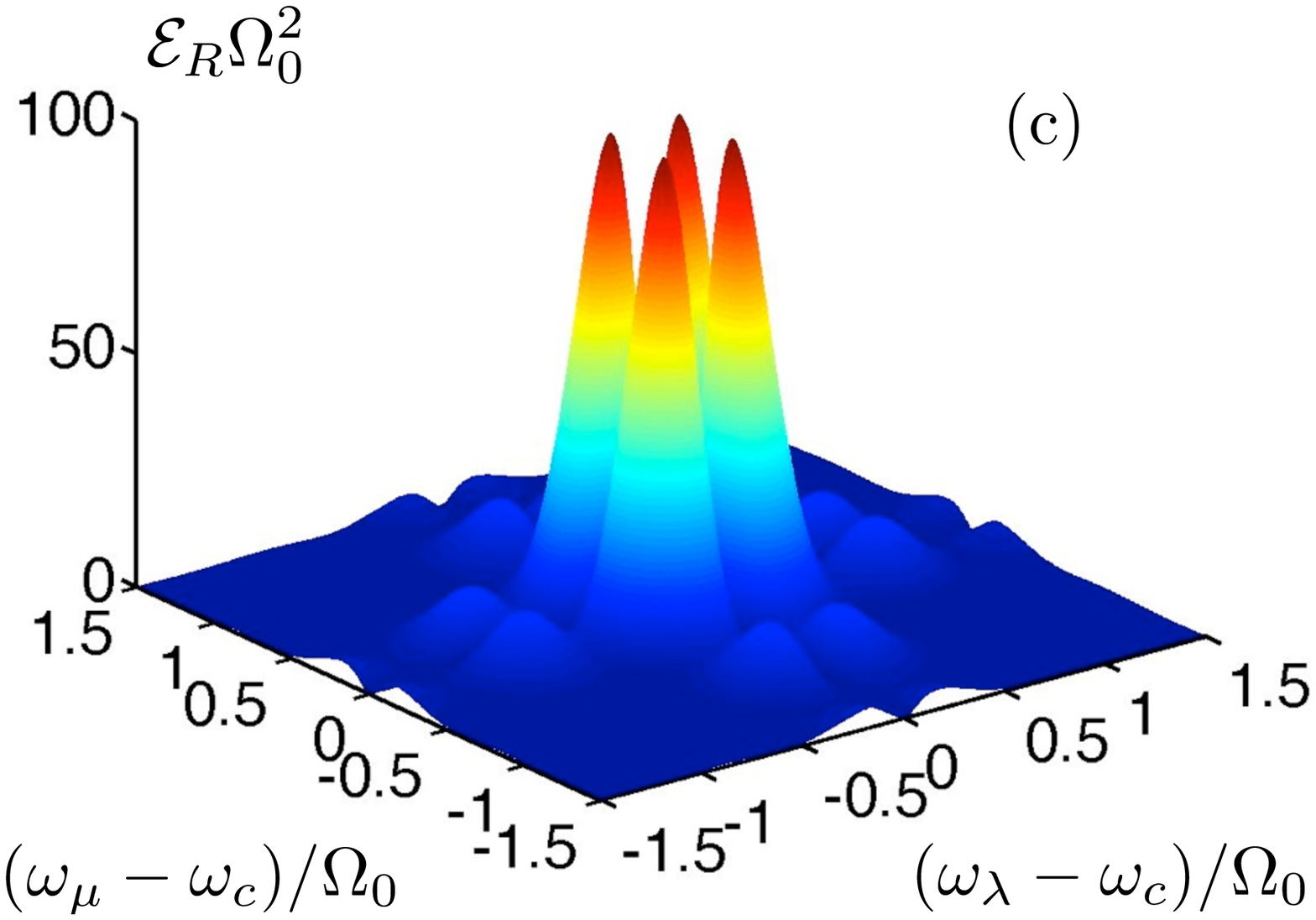}\label{fig5c}}\qquad\qquad
    \subfigure{\includegraphics[width=0.8\columnwidth]{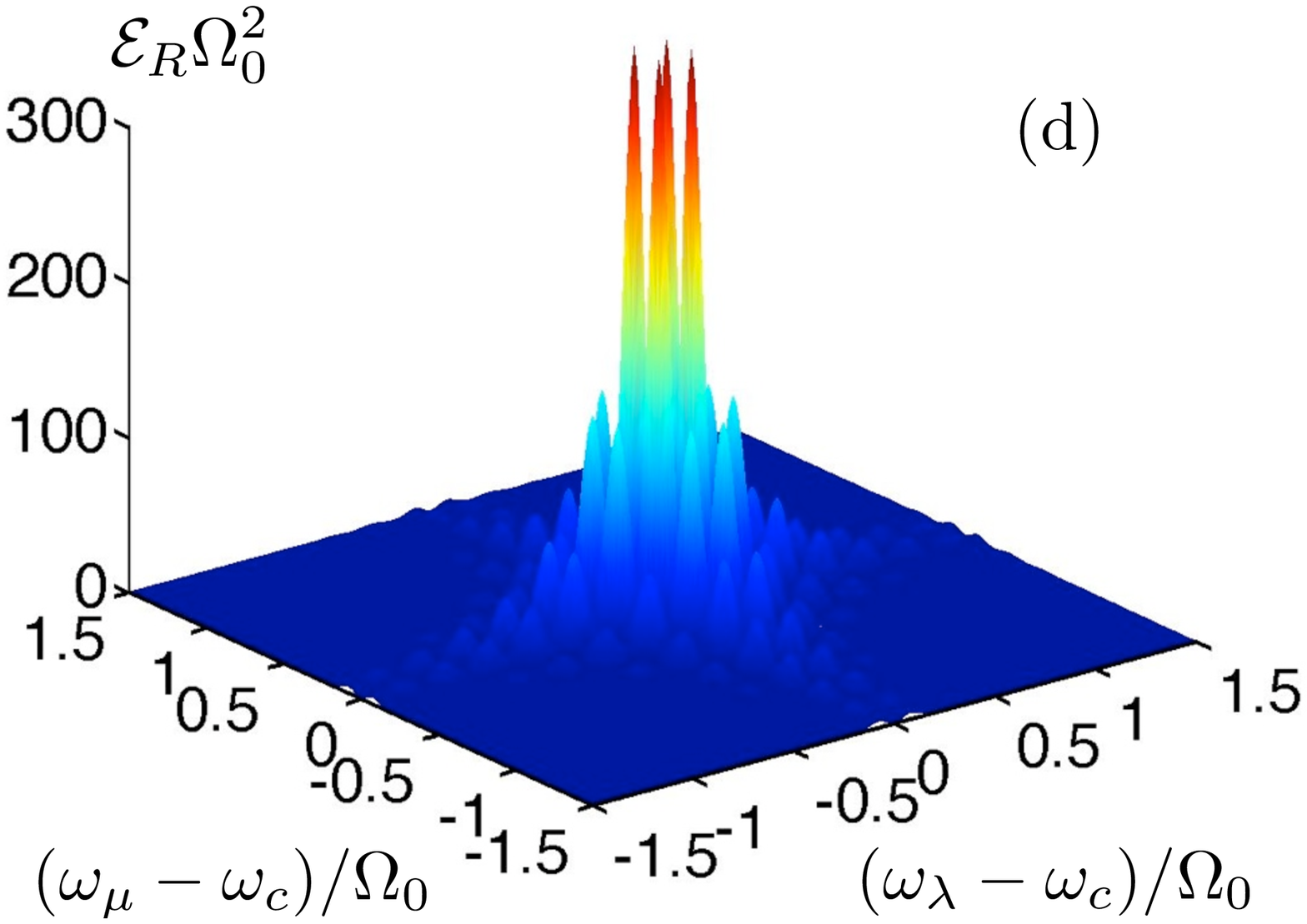}\label{fig5d}}
    \caption{(Color online) The atom-modes density of entanglement $\mathcal{E}_{A}(\omega_\lambda,t)$ as a function of time and the mode frequency $\omega_\lambda$ (a), and the density of entanglement $\mathcal{E}_{R}(\omega_\lambda,\omega_\mu,t)$ between a mode $\omega_\mu=\omega_c+0.1\Omega_0$ and the rest of the reservoir modes as a function of time (b). Figures (c) and (d) are snapshots for the mode-mode density of entanglement $\mathcal{E}_{R}(\omega_\lambda,\omega_\mu,t)$ for $\Omega_0 t=10$ and $\Omega_0 t=30$ respectively. For all figures $\Gamma_1=10\Omega_0$, $\Gamma_2=0.2\Omega_0$ and $W_1=50W_2$. }
    \label{fig5}
  \end{center}
\end{figure*}

For the strong coupling regime, as shown earlier, the dynamics are similar to those for an atom coupled to a reservoir with a Lorentzian structure function. Thus, entanglement dynamics will be similar for the atom-reservoir system under consideration in Ref.~\cite{Lazarou2011}. The main feature for the reservoir density of entanglement is a pronounced Rabi splitting. Furthermore, this splitting results in decaying oscillations in the atom-modes density of entanglement.

On the other hand, for the weak coupling regime dynamics are different. The formation of the trapping state is associated with a continuous population exchange between the atom and individual modes whilst the net flow of probability is equal to zero as discussed before.  As a consequence, both entanglement distributions change in time, as can be seen in \ref{fig5a} for $\mathcal{E}_{A}(\omega_\lambda,t)$ and in \ref{fig5b} for  $\mathcal{E}_R(\omega_\lambda,\omega_\mu,t)$. Both distributions do not reach a steady state in the long time limit. This feature for $\mathcal{E}_R(\omega_\lambda,\omega_\mu,t)$ is also evidenced in Figs. \ref{fig5c} and \ref{fig5d}, where the density of entanglement for the reservoir modes is plotted for different times.

\begin{figure}
  \begin{center}
    \includegraphics[width=0.8\columnwidth]{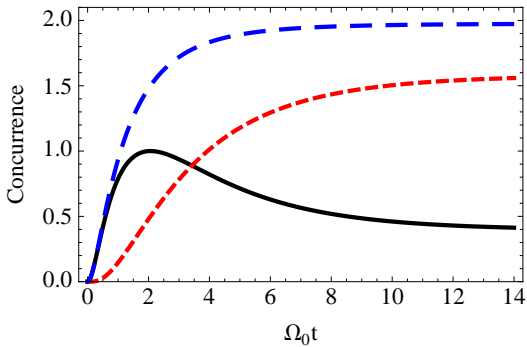}
    \caption{(Color online) The concurrence $C^2_A(t)$ (black solid), $C^2_R(t)$ (red short-dashed) and the total concurrence $C^2(t)$ (blue dashed), for $\Gamma_1=10\Omega_0$, $\Gamma_2=0.2\Omega_0$ and $W_1=50W_2$. }
    \label{fig7}
  \end{center}
\end{figure}

In contrast to this, due to population conservation and the fact that the net population exchange $Q(t)$ in the long time limit is zero, the total entanglement between the atom and the reservoir $C^2_A(t)$ is constant for $t\rightarrow\infty$. The same is true for the total entanglement for the reservoir modes $C^2_{R}(t)$ and the total entanglement $C^2(t)$. In figure \ref{fig7}, we plot $C^2_A(t)$, $C^2_R(t)$ and the total concurrence $C^2(t)$ for the weak coupling regime. 

From this we see that at very early times, a rapid build up of entanglement takes place between the atom and the reservoir. Entanglement between the reservoir modes evolves at a much slower rate. Upon reaching a maximum, atom-reservoir entanglement follows a decay reaching a steady state at about the same time as the entanglement between the reservoir modes does. This point in time corresponds to the formation of the final trapping state between the atom and the reservoir. 
\section{Conclusions}\label{conclusion}
In this work we have studied correlations and entanglement for an atom-photon bound state.  
Such states, can be formed when an atom couples to a reservoir with
a gap in its density of modes. Their main feature is that in the long time limit the system reaches 
a steady state where the initial atomic excitation energy, is shared between the atom and the 
reservoir. 

Despite the fact that for such a state no change is expected in the long time 
limit, a careful study of intra-reservoir dynamics has revealed that this is not 
the case. For a reservoir with a single frequency gap in its structure function, 
we have shown that in the long time limit, the atom exchanges energy with
individual modes, and though the net energy flow is zero, a permanent effective coupling 
between the modes is induced. As a result the reservoir spectrum changes with
time and a steady energy distribution is never reached. 

The existence of the atom-photon bound state is explained, in terms of the pseudomode description, with the formation of a dark state by the atom and one of the  pseudomodes which are both coherently coupled to that pseudomode which connects to the rest of the environment. In  general, the population trapping occurs in the weak coupling regime whereas with strong coupling and no trapping, also the dynamics within the environment begins to resemble the one obtained by single Lorentzian distribution.  Furthermore, we have studied in the detail the entanglement dynamics between the atom and the pseudomodes, and within the environmental modes, high-lighting several qualitative and quantitative differences between the weak and strong coupling regimes.

The results and methods used in this work can be extended and applied
to systems where the reservoir density of  modes has a broader gap or a more complicated 
structure. Such evolved reservoir structures can be encountered in photonic 
crystals, or when considering the problem of atom-laser outcoupling from Bose-Einstein condensates.
\acknowledgments
This work was supported by the Jenny and Antti Wihuri Foundation, Magnus Ehrnrooth Foundation, Vilho, Yrj\"o and Kalle
V\"ais\"al\"a Foundation, the Acedemy of Finland (project 259827), and the COST Action MP1006.

\appendix
\section{Solution of the Schr\"odinger equation and the pseudomodes method}\label{secA1}
Starting with an atom initially excited and the reservoir in a vacuum state, the system's wave function for $t>0$ will be 
\begin{equation}\label{eqA1}
\vert\psi(t)\rangle=c_{a}(t)\vert 1_a\rangle\vert{\bf 0}\rangle+\sum_{\lambda}c_\lambda(t)\vert 0_a\rangle\vert
\psi_\lambda\rangle.
\end{equation}
The collective vacuum state for all reservoir modes $\omega_\lambda$ is
\begin{equation}\label{eqA2}
\vert{\bf 0}\rangle=\prod_{\lambda}\vert0_\lambda\rangle,
\end{equation}
and the state with a single excitation in one of the reservoir modes is
\begin{equation}\label{eqA3}
\vert\psi_\lambda\rangle=\vert1_\lambda\rangle\prod_{k\neq\lambda}\vert0_k\rangle.
\end{equation} 
At $t=0$ we have that $c_a(0)=1$ and $c_\lambda(0)=0$.

The coefficients $c_a(t)$ and $c_\lambda(t)$ can be obtained by solving the Schr\"odinger equations
\begin{subequations}\label{eqA4}
\begin{align}
&\imu\dot{\tilde{c}}_a(t)=\sum_\lambda g_\lambda e^{-\imu\delta_\lambda t}\tilde{c}_\lambda(t), \label{A4a} \\ \nonumber \\
&\imu\dot{\tilde{c}}_\lambda(t)= g_\lambda e^{\imu\delta_\lambda t}\tilde{c}_a(t),\label{A4b} 
\end{align}
\end{subequations}
where the detuning between the atomic transition and the mode $\lambda$ is $\delta_\lambda=\omega_\lambda-\omega_0$.
The amplitudes in the interaction picture are $\tilde{c}_a(t)=e^{\imu\omega_0 t}c_a(t)$ and $\tilde{c}_\lambda(t)=e^{\imu\omega_\lambda t}c_\lambda(t)$.

To derive $c_a(t)$ and $c_\lambda(t)$ one can numerically integrate  Eqs. (\ref{A4a}) and (\ref{A4b}) using a discretization technique \cite{Lambropoulos2000,Nikolopoulos1999,Nikolopoulos2000} or by using 
the Laplace transform \cite{John1994,Kofman1994}. A different approach is that offered by the pseudomodes method \cite{Garraway1997a,Garraway1997b}. The main feature of this technique, is that the infinitely many equations for the reservoir modes can be replaced by a  finite number of equations. Thus the computational effort is substantially reduced. In addition to this,  the pseudomodes method has provided an intuitive insight into non-Markovian dynamics, which are observed when an atom strongly couples to its environment \cite{Mazzola2009}.

When the reservoir structure function is analytic with a finite number of poles in the lower complex plane, Eqs. (\ref{A4a}) and (\ref{A4b}) can be replaced 
by a set of equivalent equations \cite{Garraway1997a,Garraway1997b}. In this new set of equations the atom couples to a finite set of fictitious modes, the pseudomodes,
where each of these modes has a one--to--one correspondence to the poles of $D(\omega)$.   

For the structure function $D(\omega)$ in Eq. (\ref{eq5}), the analysis for arbitrary widths ($\Gamma_1$,$\Gamma_2$) and weights ($W_1$, $W_2$) was previously carried out, see Ref. \cite{Garraway1997b}. Here we focus only the on perfect gap case i.e. $D(\omega_c)=0$, where the equations for the atomic excitation $c_a(t)$ and the 
two pseudomodes $a_1(t)$ and $a_2(t)$ are \cite{Garraway1997b}
\begin{subequations}\label{eqA5}
\begin{align}
\imu\dot{c}_a(t)=&\omega_0 c_a(t)+\Omega_0 a_2(t), \label{A5a} \\ \nonumber \\
\imu\dot{a}_1(t)=&\omega_c a_1(t) + \frac{\sqrt{\Gamma_1\Gamma_2}}{2}a_2(t), \label{A5b}\\ \nonumber \\
\imu\dot{a}_2(t)=&\left(\omega_c-\imu\frac{\Gamma_1+\Gamma_2}{2}\right)a_2(t) \nonumber \\  \label{A5c} \\  &+\Omega_0 c_a(t)+\frac{\sqrt{\Gamma_1\Gamma_2}}{2}a_1(t) \nonumber .
  \end{align}
\end{subequations}
These equations can be associated to the following master equation
\begin{equation}\label{eqA6}
\begin{split}
\dot{\rho}(t)=&-\imu\left[H_0,\rho(t)\right]-\frac{\Gamma_1+\Gamma_2}{2}\Big(\hat{a}_2^\dagger\hat{a}_2\rho(t) \\ \\
	&-2\hat{a}_2\rho(t)\hat{a}^\dagger_2+\rho(t)\hat{a}^\dagger_2\hat{a}_2\Big),
\end{split}
\end{equation}
with the Hamiltonian
\begin{equation}\label{eqA7}
\begin{split}
H_0=&\omega_0\vert1_a\rangle\langle1_a\vert+\omega_c\left(\hat{a}^\dagger_1\hat{a}_1+\hat{a}^\dagger_2\hat{a}_2\right) \\ \\
&+\Omega_0\left(\hat{a}^\dagger_2\vert0_a\rangle\langle1_a\vert+\hat{a}_2\vert1_a\rangle\langle0_1\vert\right) \\ \\
&+\frac{\sqrt{\Gamma_1\Gamma_2}}{2}\left(\hat{a}^\dagger_1\hat{a}_2+\hat{a}_1\hat{a}^\dagger_2\right),
\end{split}
\end{equation}
where $\hat{a}_1$ $(\hat{a}^\dagger_1)$ and $\hat{a}_2$ $(\hat{a}^\dagger_2)$ are the annihilation (creation) operators for the
two pseudomodes respectively.

The solution for the master equation(\ref{eqA6}) reads 
\begin{equation}\label{eqA8}
\rho(t)=\Pi_j(t)\vert0_a0_10_2\rangle\langle0_a0_10_2\vert+\vert\tilde{\psi}(t)\rangle\langle\tilde{\psi}(t)\vert, 
\end{equation}
where 
\begin{equation}\label{eqA9}
\vert\tilde{\psi}(t)\rangle=c_a(t)\vert1_a0_10_2\rangle+a_1(t)\vert0_a1_10_2\rangle+a_2(t)\vert0_a0_11_2\rangle.
\end{equation}
The vacuum state population $\Pi_j(t)$ is given by
\begin{equation}\label{eqA10}
\Pi_j(t)=\frac{\Gamma_1+\Gamma_2}{2}\int_{0}^{t}\textrm{d}\tau\vert a_2(\tau)\vert^2.
\end{equation}
The Fock states with zero or one excitation for the two pseudomodes are $\vert 0_1\rangle$ $(\vert0_2\rangle)$ and $\vert1_1\rangle$ ($\vert1_2\rangle)$ respectively. From Eq. (\ref{eqA8}) we see that the atom and the pseudomodes are in a mixed state.

Equations (\ref{A5a})-(\ref{A5c}) are linear with time-independent coefficients and 
solutions can be easily obtained with the Laplace transform method. With the initial population
for the atom being $c_a(0)=1$, and both pseudomodes in a vacuum state, $a_1(0)=a_2(0)=0$, we get for $c_a(t)$ 
\begin{equation}\label{eqA11}
\begin{split}
c_a(t)=&\frac{4e^{\imu\omega_0 t}}{4\Gamma^2+\Omega^2}\Bigg[\frac{\Gamma_1\Gamma_2}{4}+\frac{2\Omega^2_0}{\Omega}e^{-\Gamma t}\Bigg(\Gamma\sin\left(\frac{\Omega t}{2}\right) \\ \\
&+\frac{\Omega}{2}\cos\left(\frac{\Omega t}{2}\right)\Bigg)\Bigg],
\end{split}
\end{equation}
and for the pseudomodes 
\begin{equation}\label{eqA12}
\begin{split}
a_1(t)=&-\frac{2\sqrt{\Gamma_1\Gamma_2}\Omega_0e^{\imu\omega_0 t}}{(4\Gamma^2+\Omega^2)}\Bigg[1-e^{-\Gamma t}\Bigg(\cos\left(\frac{\Omega t}{2}\right) \\ \\ &+\frac{2\Gamma}{\Omega}\sin\left(\frac{\Omega t}{2}\right)\Bigg)\Bigg],
\end{split}
\end{equation}
and
\begin{equation}\label{eqA13}
\begin{split}
a_2(t)=-\frac{2\imu\Omega_0e^{\imu\omega_0 t}}{\Omega}\sin\left(\frac{\Omega t}{2}\right)e^{-\Gamma t}.
\end{split}
\end{equation}
Here we consider only the resonant case $\omega_0=\omega_c$. The decay rate $\Gamma$ and the Rabi frequency $\Omega$ are
\begin{equation}\label{eqA14}
\Gamma=\frac{\Gamma_1+\Gamma_2}{4},
\end{equation}
and
\begin{equation}\label{eqA15}
\Omega=\frac{1}{2}\sqrt{16\Omega^2_0-(\Gamma_1-\Gamma_2)^2}.
\end{equation}

Substituting (\ref{eqA12}) in (\ref{eqA10}) we get for the vacuum state population 
\begin{equation}\label{eqA16}
\begin{split}
\Pi_j(t)=&\frac{16\Gamma\Omega^2_0}{\Omega^2}\Bigg[\frac{\Omega^2}{4\Gamma(4\Gamma^2+\Omega^2)}-\frac{e^{-2\Gamma t}}{4\Gamma} \\ \\ &+\frac{\Gamma\cos(\Omega t)-\frac{\Omega}{2}\sin(\Omega t)}{(4\Gamma^2+\Omega^2)}e^{-2\Gamma t}\Bigg].
\end{split}
\end{equation}
Finally using Eqs. (\ref{eqA1}) and (\ref{A4b}) we get the amplitudes $c_\lambda(t)$ for the reservoir modes
\begin{equation}\label{eqA17}
\begin{split}
c_\lambda(t)=&-\frac{4\imu e^{\imu\omega_\lambda t}g_\lambda}{4\Gamma^2+\Omega^2}\Bigg[\frac{\Gamma_1\Gamma_2}{2\delta_\lambda}e^{\imu\delta_\lambda t/2}\sin\left(\frac{\delta_\lambda t}{2}\right) \\ \\ &+\frac{4\Omega^2_0\left(2\Gamma-\imu\delta_\lambda\right)}{4(\Gamma-\imu\delta_\lambda)^2+\Omega^2}\Bigg(1-e^{\imu\delta_\lambda t-\Gamma t}\cos\left(\frac{\Omega t}{2}\right)\Bigg) \\  \\&+\frac{2\Omega^2_0(4(\imu\delta_\lambda\Gamma-\Gamma^2)+\Omega^2)}{\Omega(4(\Gamma-\imu\delta_\lambda)^2+\Omega^2)}e^{\imu\delta_\lambda t-\Gamma t}\sin\left(\frac{\Omega t}{2}\right)\Bigg],
\end{split}
\end{equation}
where $\delta_\lambda=\omega_\lambda-\omega_c$ is the detuning between the $\lambda$ mode and the gap frequency $\omega_c$.
\bibliography{CLazarou.bbl}
\end{document}